# The Gluon Exchange Model in proton-nucleus collisions

Marek Jeżabek[1] and Andrzej Rybicki[1]

[1] Institute of Nuclear Physics, Polish Academy of Sciences,
Radzikowskiego 152, 31-342 Kraków, Poland

**Abstract**

We apply our recently formulated Gluon Exchange Model (GEM) to baryon production in proton-nucleus reactions involving $N>1$ proton-nucleon collisions. We propose a description scheme for the process of soft color octet (gluon) exchange, based on the assumption that probabilities to form an effective diquark are equal for all allowed pairs of quarks. The latter effective diquark can form either from two valence, one valence and one sea, or from two sea quarks. Consequently we calculate the probabilities for different color configurations involving diquarks of valence-valence, valence-sea and sea-sea type. These probabilities appear to depend on the number of exchanged gluons, which results in increasing baryon stopping as a function of the number of proton-nucleon collisions in the nucleus. As such, the nuclear stopping power appears to be governed by the emergence of new color configurations as a function of $N$ rather than by the energy loss of the original valence diquark.
The advantage of our approach lies in its high predictive power which makes it verifiable by the new, precise data on proton and neutron production from the CERN SPS. The latter verification, and a set of predictions for the $N$-dependence of the baryon stopping process, are included in the letter.

## 1. Introduction

In a recent letter [1] we proposed a new model for diffractive and inelastic hadron-induced collisions, based on the partonic (constituent) structure of the incoming hadrons and on the exchange of soft color octets (gluons) between the constituents. With this Gluon Exchange Model (GEM), we achieved a precise description of proton and neutron spectra in the projectile hemisphere of pp reactions at $\sqrt{s_{NN}}$=17.3 GeV. Presently we apply the same model to proton-nucleus reactions with the aim of obtaining better insight into the nature of the baryon stopping process (nuclear stopping power[1]). Within a general GEM scheme we study the consequences of a very restrictive, although natural statistical assumption on the color structure of quark constituents, that is both *valence* and *sea* quarks in the projectile for the process of multiple proton-nucleon collisions. This process results in the formation of well defined color configurations as a function of the number of exchanged gluons N, which builds up baryon stopping as a function of centrality and/or nuclear size. The comparison of our GEM calculations to wide coverage experimental data on proton and neutron emission from the NA49 experiment at the CERN SPS [2,3] is included in the paper. Predictions for the nuclear stopping power as a function of the number of exchanged gluons are presented for N=2, 3, 4, and 6.

    This work inherits the entire formalism of GEM as we presented in Ref. [1] (a more technical description can be found in an auxiliary paper [4]). Here we only remind that GEM can be considered as a generalization of the original Dual Parton Model [5,6] with a more complete treatment of the color quantum number and a significant, even if natural, extension of the Fock space of states available to

---

[1] This term is taken from Ref. [14].

participating protons and nucleons. As it was stated in Ref. [1], the GEM model brings the advantage of providing a complete description of the entire proton and neutron spectrum in pp collisions, including the proton "diffractive peak" at high $x_F$ which is explained as a specific case of color octet exchange between the valence quarks from one and the virtual sea quark-sea antiquark pair from the other proton. As it will become apparent in the course of this letter, the model brings very little freedom for the transition between pp and proton-nucleus collisions which adds up to the reliability of comparison with experimental data.

**2. Insights into the nature of nuclear stopping power**

In this section we enumerate our earlier findings which directly inspired the present analysis. These were:

(i) the qualitative difference between the shape of proton $x_F$ spectra in single proton-nucleon (pp) reactions and multiple proton-nucleon (pA) collisions (Ref. [1], Fig. 2 therein) suggesting the presence of qualitatively new processes in the latter;

(ii) the failure of our model calculation, arbitrarily limited to the valence diquark-preserving diagrams of GEM (Ref. [1], Fig. 4 therein), to explain the experimental data on pC reactions showing that the energy loss by the valence diquark alone cannot be held responsible for the dominant part of baryon stopping in pA collisions;

(iii) the result of our ad hoc fit of specific GEM diagrams to the latter experimental data (Ref. [4], Fig. 14 therein), suggesting a very important role of diagrams involving sea quarks in the transport of baryon number.

These findings brought us to a specific realization of the GEM model, relying on the concept of effective diquarks as (mostly) responsible for baryon production. The main assumption underlying this work is that the latter diquarks need to be formed from both *valence* and *sea* quarks in all the three possible combinations (valence-valence, valence-sea, sea-sea). This assumption we quantify in the following two options for the multiple proton-nucleon collision process:

(a) a single soft gluon exchange brings the valence quarks of the projectile proton into the color octet state, while the remaining N-1 gluons coupling to sea quarks remain essentially independent (the $R_8^{N-1}$ representation, see Sec. 4 for more details);

(b) two soft gluons bring the valence quarks of the projectile proton into the symmetric color decuplet state, while the remaining N-2 gluons coupling to sea quarks remain independent (the $R_{10}^{N-2}$ representation).

A more detailed discussion will be presented in the Secs. 3-5 below.

**3. Irreducible representations of SU$_3$**

Irreducible representations (IRs) of SU$_3$,

$$R_{m_1,m_2,m_3} := (m_1, m_2, m_3)$$

are specified by triplets of integers

$$m_1 \geq m_2 \geq m_3 \geq 0.$$

The dimension of the representation ( $m_1$, $m_2$ , $m_3$ ) is:

$$D(m_1, m_2, m_3) = [m_1 - m_3, m_2 - m_3]$$

where

$$[k, l] := \frac{1}{2}(k - l + 1)(k + 2)(l + 1).$$

The bases of these representations can be presented graphically as Young tableaux with three rows of length $m_1$, $m_2$ and $m_3$, respectively. In the following we use also the dimensions of some IRs instead of ( $m_1$ , $m_2$ , $m_3$ ).

**3.1 Examples of IRs**

The following basic IRs will be essential for the discussion made in this letter.

| | | |
|---|---|---|
| Fundamental representation (triplet): | (1,0,0) | or $\underset{\sim}{3}$ |
| Anti-triplet (complex conjugate): | (1,1,0) | or $\underset{\sim}{3^*}$ |
| Singlet: | (1,1,1) | or $\underset{\sim}{1}$ |
| Adjoint (octet): | (2,1,0) | or $\underset{\sim}{8}$ |
| Decuplet: | (3,0,0) | or $\underset{\sim}{10}$ . |

**3.2 Notation**

In general the tensor product of representations $R_a$ and $R_b$ is denoted as $R_a \otimes R_b$ . However, a shortened notation is used for

$$\prod_{\otimes}^{N} \underset{\sim}{3} := \underbrace{\underset{\sim}{3} \otimes \underset{\sim}{3} \otimes \ldots \underset{\sim}{3}}_{N\ times} := \underset{\sim}{3}^N$$

and the decomposition of a reducible representation into a direct sum of IRs is written as

$$R = \sum_{\oplus} M_{m_1, m_2, m_3}\, (m_1, m_2, m_3)$$

where $M_{m_1,m_2,m_3}$ denotes the multiplicity of the representation ( $m_1$ , $m_2$ , $m_3$ ) in the direct sum.

## 4. Color configurations in pA collisions

In the framework of GEM the production of secondary particles in pA inelastic collisions is governed by the color state of constituents of the projectile (p) and the target (A). Therefore it is essential to consider the color configuration of valence and sea quarks in the proton, after the exchange of N gluons (color octets) with N nucleons in the nucleus. For such a multiple scattering process, the scattered proton is described by a state in the Fock space including valence quarks as well as sea quark-antiquark pairs [1]. The first and dominant contribution to this process arises when one of the gluons couples to the valence quarks and the remaining N − 1 octets are exchanged with sea quark – antiquark pairs. The constituent quarks in the proton, both valence and sea ones, are in the color representation

$$R_8^{N-1} = \underset{\sim}{8} \otimes \underset{\sim}{3}^{N-1},$$

which can be expressed as a direct sum of IRs ( $m_1$ , $m_2$ , $m_3$ ), such that

$$m_1 + m_2 + m_3 = N + 2 \ .$$

The second possibility which we consider in this letter is that the valence quarks in the proton absorb two color octets and are in a color symmetric decuplet state. The remaining N − 2 gluons couple to quark – antiquark sea constituents of the proton, and the color configuration of all quark constituents in the proton is

$$R_{10}^{N-2} = \underset{\sim}{10} \otimes \underset{\sim}{3}^{N-2},$$

which can again be expressed as a direct sum of IRs ( $m_1$ , $m_2$ , $m_3$ ), such that

$$m_1 + m_2 + m_3 = N + 1 \ .$$

As it was already fully apparent in our earlier work [1], the production of secondary baryons in the hemisphere of the projectile depends in a most crucial way on the *diquark*, i.e. a system of two constituent quarks in the proton. The diquark must be in a *color antitriplet* state which implies that it can be built from two constituent quarks which are not color symmetric. From a technical point of view these quarks must be placed in different rows of the corresponding Young tableau for a given IR of color $SU_3$.

In this letter we study the consequences of the relatively simple assumption that exclusion of quark pairs symmetric in color is the only condition imposed on diquark formation, that is, that probabilities to form a diquark are equal for all the allowed pairs of quarks, independently on whether they were valence or sea quarks in the initial state of the collision. We underline the corresponding effective character of the diquark which goes well in line with our earlier findings [1,4,7,8].

The allowed types of diquarks depend on the color states of constituent quarks:

**(a)** for the representations $R_8^{N-1}$, where N ≥ 1, the diquarks may be composed of two valence quarks (*i.e.*, be of VV type), of one valence and one sea quark (VS type), or of two sea quarks (SS type). For N = 1, however, only the VV type is allowed, and the last option (SS) is possible only for N ≥ 3 ;

**(b)** for the representations $R_{10}^{N-2}$, where $N \geq 2$ the types VS and SS are allowed, but VV is not. An important new option is that all quarks are in a fully symmetric color state and therefore *no diquark can be built* out of the constituents of the proton (the 0 type). In particular for $N = 2$, the corresponding representation (decuplet) is fully symmetric and the probability $P_0\binom{10}{-}$ of no diquark is equal to unity. In general for $N \geq 2$, the probability for the 0 type is

$$P_0(R_{10}^{N-2}) = \frac{(N+2)(N+3)}{20 \cdot 3^{N-2}} \ .$$

Presently for a given IR ( $m_1$ , $m_2$ , $m_3$ ), we define a number $\Sigma(\, m_1\, ,\, m_2\, ,\, m_3\, )$ :

$$\Sigma(m_1, m_2, m_3) = m_1 \cdot m_2 + m_2 \cdot m_3 + m_3 \cdot m_1$$

which is equal to the number of quark pairs allowed to form a diquark, *i.e.*, these pairs which are composed of quarks in different rows of the corresponding Young tableaux. We define also the functions $n_{VV}(\, m_1\, ,\, m_2\, ,\, m_3\, )$, $n_{VS}(\, m_1\, ,\, m_2\, ,\, m_3\, )$, and $n_{SS}(\, m_1\, ,\, m_2\, ,\, m_3\, )$ which are equal to the numbers of allowed quark pairs composed of two valence, one valence and one sea, and two sea quarks respectively.

It follows that

$$\Sigma(m_1, m_2, m_3) = n_{VV}(m_1, m_2, m_3) + n_{VS}(m_1, m_2, m_3) + n_{SS}(m_1, m_2, m_3)$$

and

$n_{VV}(m_1, m_2, m_3) = 2$, $n_{VS}(m_1, m_2, m_3) = m_1 + 2\,m_2 + 3\,m_3 - 4$ for the case **(a)** above,

$n_{VV}(m_1, m_2, m_3) = 0$, $n_{VS}(m_1, m_2, m_3) = 3(m_2 + m_3)$ for the case **(b)** above.

As it follows from our simple assumption, for the two families of reducible representations $R_8^{N-1}$ and $R_{10}^{N-2}$, the probability distributions of the three types of diquarks (VV, VS, SS) and of no diquark are given by the following formulae:

* *For the case (a), with $N \geq 1$ :*

$$P_A(R_8^{N-1}) = \frac{1}{8 \cdot 3^{N-1}} \sum_{m_1+m_2+m_3=N+2} M_{m_1,m_2,m_3}\, D(m_1, m_2, m_3)\, \frac{n_A(m_1, m_2, m_3)}{\Sigma(m_1, m_2, m_3)} \qquad (1)$$

where $A$ = VV, VS and SS, and the numbers $M_{m_1,m_2,m_3}$ denote the non-zero multiplicities in the direct sum of irreducible representations

$$R_8^{N-1} = \sum_\oplus M_{m_1,m_2,m_3}\, (m_1, m_2, m_3)\ .$$

In the above equality the dimensions of representations on both sides must be equal, which means that

$$D(R_8^{N-1}) = 8 \cdot 3^{N-1} = \sum_{m_1+m_2+m_3=N+2} M_{m_1,m_2,m_3} D(m_1, m_2, m_3) \ . \tag{2}$$

It follows that

$$P_{VV}(R_8^{N-1}) + P_{VS}(R_8^{N-1}) + P_{SS}(R_8^{N-1}) = 1 \ .$$

*For the case (b), with N ≥ 2 :*

$$P_{VV}(R_{10}^{N-2}) = 0 \ ,$$

$$P_A(R_{10}^{N-2}) = \frac{1}{10 \cdot 3^{N-2}} \sum_{\substack{m_1+m_2+m_3=N+1 \\ m_2 \neq 0}} M_{m_1,m_2,m_3} D(m_1, m_2, m_3) \frac{n_A(m_1, m_2, m_3)}{\Sigma(m_1, m_2, m_3)} \tag{3}$$

where $A$ = $VS$ and $SS$, and the numbers $M_{m_1,m_2,m_3}$ denote the non-zero multiplicities in the direct sum of irreducible representations

$$R_{10}^{N-2} = \Sigma_\oplus M_{m_1,m_2,m_3} \ (m_1, m_2, m_3) \ .$$

A common dimension of both reducible representations in the above equality is

$$D(R_{10}^{N-2}) = 10 \cdot 3^{N-2} = \sum_{m_1+m_2+m_3=N+1} M_{m_1,m_2,m_3} D(m_1, m_2, m_3) \tag{4}$$

A fully symmetric representation *(N+1, 0, 0)* in the above direct sum appears only once, and thus $M_{N+1,0,0}$ = 1 . Its dimension *D(N+1,0,0)* is equal to

$$[N + 1, 0] = \tfrac{1}{2}(N + 2)(N + 3) \ .$$

A simple calculation gives

$$\sum_{\substack{m_1+m_2+m_3=N+1 \\ m_2 \neq 0}} M_{m_1,m_2,m_3} D(m_1, m_2, m_3) = D(R_{10}^{N-2}) \left[1 - P_0(R_{10}^{N-2})\right] \tag{5}$$

and

$$P_{VS}(R_{10}^{N-2}) + P_{SS}(R_{10}^{N-2}) + P_0(R_{10}^{N-2}) = 1 \ .$$

## 5. Explicit formulae for representations

### 5.1 Representations $R_8^{N-1}$

In this section explicit formulae are presented for the family of representations $R_8^{N-1}$. These are expressed as direct sums of IRs, for $1 \leq N \leq 9$ soft gluons exchanged between the projectile proton and N nucleons in the nucleus.

$R_8^0 = \underset{\sim}{8} = (2,1,0)$

$R_8^1 = \underset{\sim}{8} \otimes \underset{\sim}{3} = (3,1,0) \oplus (2,2,0) \oplus (2,1,1)$

$R_8^2 = \underset{\sim}{8} \otimes \underset{\sim}{3}^2 = (4,1,0) \oplus 2 \cdot (3,2,0) \oplus 2 \cdot (3,1,1) \oplus 2 \cdot (2,2,1)$

$R_8^3 = (5,1,0) \oplus 3 \cdot (4,2,0) \oplus 3 \cdot (4,1,1) \oplus 2 \cdot (3,3,0) \oplus 6 \cdot (3,2,1) \oplus 2 \cdot (2,2,2)$

$R_8^4 = (6,1,0) \oplus 4 \cdot (5,2,0) \oplus 4 \cdot (5,1,1) \oplus 5 \cdot (4,3,0) \oplus 12 \cdot (4,2,1) \oplus 8 \cdot (3,3,1) \oplus 8 \cdot (3,2,2)$

$R_8^5 = (7,1,0) \oplus 5 \cdot (6,2,0) \oplus 5 \cdot (6,1,1) \oplus 9 \cdot (5,3,0) \oplus 20 \cdot (5,2,1) \oplus 5 \cdot (4,4,0) \oplus 25 \cdot (4,3,1) \oplus 20 \cdot (4,2,2) \oplus 16 \cdot (3,3,2)$

$R_8^6 = (8,1,0) \oplus 6 \cdot (7,2,0) \oplus 6 \cdot (7,1,1) \oplus 14 \cdot (6,3,0) \oplus 30 \cdot (6,2,1) \oplus 14 \cdot (5,4,0) \oplus 54 \cdot (5,3,1) \oplus 40 \cdot (5,2,2) \oplus 30 \cdot (4,4,1) \oplus 61 \cdot (4,3,2) \oplus 16 \cdot (3,3,3)$

$R_8^7 = (9,1,0) \oplus 7 \cdot (8,2,0) \oplus 7 \cdot (8,1,1) \oplus 20 \cdot (7,3,0) \oplus 42 \cdot (7,2,1) \oplus 28 \cdot (6,4,0) \oplus 98 \cdot (6,3,1) \oplus 70 \cdot (6,2,2) \oplus 14 \cdot (5,5,0) \oplus 98 \cdot (5,4,1) \oplus 155 \cdot (5,3,2) \oplus 91 \cdot (4,4,2) \oplus 77 \cdot (4,3,3)$

$R_8^8 = (10,1,0) \oplus 8 \cdot (9,2,0) \oplus 8 \cdot (9,1,1) \oplus 27 \cdot (8,3,0) \oplus 56 \cdot (8,2,1) \oplus 48 \cdot (7,4,0) \oplus 160 \cdot (7,3,1) \oplus 112 \cdot (7,2,2) \oplus 42 \cdot (6,5,0) \oplus 224 \cdot (6,4,1) \oplus 323 \cdot (6,3,2) \oplus 112 \cdot (5,5,1) \oplus 344 \cdot (5,4,2) \oplus 232 \cdot (5,3,3) \oplus 168 \cdot (4,4,3)$

We note that Eq. (2) provides a useful cross check of these formulae:

$D(R_8^1) = 24 = 15 + 6 + 3$

$D(R_8^2) = 72 = 24 + 2 \cdot 15 + 2 \cdot 6 + 2 \cdot 3$

$D(R_8^3) = 216 = 35 + 3 \cdot 27 + 3 \cdot 10 + 2 \cdot 10 + 6 \cdot 8 + 2 \cdot 1$

$D(R_8^4) = 648 = 48 + 4 \cdot 42 + 4 \cdot 15 + 5 \cdot 24 + 12 \cdot 15 + 8 \cdot 6 + 8 \cdot 3$

$D(R_8^5) = 1944 = 63 + 5 \cdot 60 + 5 \cdot 21 + 9 \cdot 42 + 20 \cdot 24 + 5 \cdot 15 + 25 \cdot 15 + 20 \cdot 6 + 16 \cdot 3$

$$D(R_8^6) = 5832 = 80 + 6 \cdot 81 + 6 \cdot 28 + 14 \cdot 64 + 30 \cdot 35 + 14 \cdot 35 + 54 \cdot 27 + \\ 40 \cdot 10 + 30 \cdot 10 + 61 \cdot 8 + 16 \cdot 1$$

$$D(R_8^7) = 17496 = 99 + 7 \cdot 105 + 7 \cdot 36 + 20 \cdot 90 + 42 \cdot 48 + 28 \cdot 60 + 98 \cdot 42 + \\ 70 \cdot 15 + 14 \cdot 21 + 98 \cdot 24 + 155 \cdot 15 + 91 \cdot 6 + 77 \cdot 3$$

$$D(R_8^8) = 52488 = 120 + 8 \cdot 132 + 8 \cdot 45 + 27 \cdot 120 + 56 \cdot 63 + 48 \cdot 90 + 160 \cdot 60 + \\ 112 \cdot 21 + 42 \cdot 48 + 224 \cdot 42 + 323 \cdot 24 + 112 \cdot 15 + 344 \cdot 15 + 232 \cdot 6 + \\ 168 \cdot 3$$

## 5.2 Representations $R_{10}^{N-2}$

In this section we present explicit formulae for the family of representations $R_{10}^{N-2}$. These are expressed as direct sums of IRs, for $2 \le N \le 9$ soft gluons.

$$R_{10}^0 = \underset{\sim}{10} = (3,0,0)$$

$$R_{10}^1 = \underset{\sim}{10} \otimes \underset{\sim}{3} = (4,0,0) \oplus (3,1,0)$$

$$R_{10}^2 = \underset{\sim}{10} \otimes \underset{\sim}{3}^2 = (5,0,0) \oplus 2 \cdot (4,1,0) \oplus (3,2,0) \oplus (3,1,1)$$

$$R_{10}^3 = (6,0,0) \oplus 3 \cdot (5,1,0) \oplus 3 \cdot (4,2,0) \oplus 3 \cdot (4,1,1) \oplus (3,3,0) \oplus 2 \cdot (3,2,1)$$

$$R_{10}^4 = (7,0,0) \oplus 4 \cdot (6,1,0) \oplus 6 \cdot (5,2,0) \oplus 6 \cdot (5,1,1) \oplus 4 \cdot (4,3,0) \oplus 8 \cdot (4,2,1) \oplus \\ 3 \cdot (3,3,1) \oplus 2 \cdot (3,2,2)$$

$$R_{10}^5 = (8,0,0) \oplus 5 \cdot (7,1,0) \oplus 10 \cdot (6,2,0) \oplus 10 \cdot (6,1,1) \oplus 10 \cdot (5,3,0) \oplus \\ 20 \cdot (5,2,1) \oplus 4 \cdot (4,4,0) \oplus 15 \cdot (4,3,1) \oplus 10 \cdot (4,2,2) \oplus 5 \cdot (3,3,2)$$

$$R_{10}^6 = (9,0,0) \oplus 6 \cdot (8,1,0) \oplus 15 \cdot (7,2,0) \oplus 15 \cdot (7,1,1) \oplus 20 \cdot (6,3,0) \oplus \\ 40 \cdot (6,2,1) \oplus 14 \cdot (5,4,0) \oplus 45 \cdot (5,3,1) \oplus 30 \cdot (5,2,2) \oplus 19 \cdot (4,4,1) \oplus \\ 30 \cdot (4,3,2) \oplus 5 \cdot (3,3,3)$$

$$R_{10}^7 = (10,0,0) \oplus 7 \cdot (9,1,0) \oplus 21 \cdot (8,2,0) \oplus 21 \cdot (8,1,1) \oplus 35 \cdot (7,3,0) \oplus \\ 70 \cdot (7,2,1) \oplus 34 \cdot (6,4,0) \oplus 105 \cdot (6,3,1) \oplus 70 \cdot (6,2,2) \oplus 14 \cdot (5,5,0) \oplus \\ 78 \cdot (5,4,1) \oplus 105 \cdot (5,3,2) \oplus 49 \cdot (4,4,2) \oplus 35 \cdot (4,3,3)$$

One can check that Eq. (4) is fulfilled:

$$D(R_{10}^1) = 30 = 15 + 15$$

$$D(R_{10}^2) = 90 = 21 + 2 \cdot 24 + 15 + 6$$

$$D(R_{10}^3) = 270 = 28 + 3 \cdot 35 + 3 \cdot 27 + 3 \cdot 10 + 10 + 2 \cdot 8$$

$$D(R_{10}^4) = 810 = 36 + 4 \cdot 48 + 6 \cdot 42 + 6 \cdot 15 + 4 \cdot 24 + 8 \cdot 15 + 3 \cdot 6 + 2 \cdot 3$$

$$D(R_{10}^5) = 2430 = 45 + 5 \cdot 63 + 10 \cdot 60 + 10 \cdot 21 + 10 \cdot 42 + 20 \cdot 24 + 4 \cdot 15 + 15 \cdot 15 + 10 \cdot 6 + 5 \cdot 3$$

$$D(R_{10}^6) = 7290 = 55 + 6 \cdot 80 + 15 \cdot 81 + 15 \cdot 28 + 20 \cdot 64 + 40 \cdot 35 + 14 \cdot 35 + 45 \cdot 27 + 30 \cdot 10 + 19 \cdot 10 + 30 \cdot 8 + 5 \cdot 1$$

$$D(R_{10}^7) = 21870 = 66 + 7 \cdot 99 + 21 \cdot 105 + 21 \cdot 36 + 35 \cdot 90 + 70 \cdot 48 + 34 \cdot 60 + 105 \cdot 42 + 70 \cdot 15 + 14 \cdot 21 + 78 \cdot 24 + 105 \cdot 15 + 49 \cdot 6 + 35 \cdot 3$$

## 6. Results

In Table 1 we present the final probability distributions for the two families of reducible representations $R_8^{N-1}$ and $R_{10}^{N-2}$, for the three types of diquarks (VV, VS, SS) and for no diquark (the 0 type). For simplicity, the two physical cases corresponding to these two families (Sec. 4) will be labeled "color octet exchange" and "color decuplet exchange" from now on.

| | $\underline{8} \otimes \underline{3}^{N-1}$ | | | $\underline{10} \otimes \underline{3}^{N-2}$ | | |
|---|---|---|---|---|---|---|
| N | V V | V S | S S | 0 | V S | S S |
| 1 | 1 | - | - | - | - | - |
| 2 | 0.5917 | 0.4083 | - | 1 | - | - |
| 3 | 0.3740 | 0.5223 | 0.1037 | 0.5 | 0.5 | - |
| 4 | 0.2520 | 0.5407 | 0.2073 | 0.2333 | 0.6238 | 0.1429 |
| 5 | 0.1784 | 0.5213 | 0.3002 | 0.1037 | 0.6179 | 0.2784 |
| 6 | 0.1319 | 0.4908 | 0.3773 | 0.0444 | 0.5733 | 0.3823 |
| 7 | 0.1010 | 0.4582 | 0.4408 | 0.0185 | 0.5234 | 0.4581 |
| 8 | 0.0797 | 0.4272 | 0.4931 | 0.0075 | 0.4770 | 0.5155 |
| 9 | 0.0644 | 0.3989 | 0.5367 | 0.0030 | 0.4366 | 0.5604 |

Table 1. Probabilities for the two families of representations as a function of N. The probabilities are computed through Eqs. (1), (3) and (5).

What is immediately apparent in the Table for the $R_8^{N-1}$ representation family as a function of N, is the rapid decrease of probability for diquark formation from two valence quarks (the VV type), and the corresponding emergence of the SS diquark type. For the color decuplet case (the $R_{10}^{N-2}$ family) it is the even more rapid suppression of the 0 type (no diquark) as a function of N. This implies that only pA reactions involving small nuclei (like pC collisions) or very specific centrality selections are suitable to study the role of this specific contribution to baryon spectra in the final state.

## 7. Comparison to experimental data

Presently the results of calculations performed in Secs. 4-6 will be put in comparison with experimental data on proton-carbon reactions at $\sqrt{s_{NN}}$=17.3 GeV. These data from the NA49 experiment [2,3] are the same we described in Ref. [1] where our model calculation, arbitrarily limited only to diagrams preserving the valence diquark, failed to describe the baryon distributions in the final state. The model calculation of the different contributions to baryon spectra from color octet ( $R_8^{N-1}$ ) and color decuplet ( $R_{10}^{N-2}$ ) exchanges is performed in a way identical to that described in Refs. [1,4]. The diagrams for color singlet fragmentation for the configurations of types VV, VS and 0 are the diagrams (c), (f), and (e) published previously in Ref. [1] (Fig. 1 therein) while the corresponding diagram for the SS type is illustrated in Fig. 1 below. The fragmentation and isospin flip functions as well as the small corrections for the presence of diagrams with single diffraction in the nucleus and for isospin differences between pp and proton-nucleon collisions are all realized identically as described in Refs. [1,4]. The only "free" parameter we take in the model is the relative weight of color octet to color decuplet contributions. As it will become apparent below the latter is strongly constrained by the experimental data.

Fig. 2 presents the distribution of net protons and net neutrons in pC reactions in which the projectile proton collides with multiple (N>1) nucleons, put together with the results of our present GEM calculation. We find that the relative weights of 94.5% color octet and 5.5% color decuplet exchange result in the best description of the data. This description we find satisfactory although we note that the present calculation still appears to underestimate the (very strong) isospin effects between proton and neutron spectra, in particular at high rapidity. This issue we find interesting and leave it for a future paper [15].

The break-up of the total proton and neutron distribution into the different contributions from Table 1 is also included in the Figure. For the strongly dominant color octet exchange an evident hierarchy is apparent as a function of rapidity. The valence-valence (VV) diquark appears mostly responsible for the forward rapidity and the valence-sea (VS) diquark for the intermediate rapidity part of the baryon spectrum, while the contribution from the sea-sea (SS) diquark is localized closest to central rapidity ($y=0$). The very small weight obtained for color decuplet exchange evidently precludes a precise delimitation of this contribution; we are rather inclined to conclude that an upper limit of the order of ~10% for color decuplet exchange is suggested by the experimental data, building up, together with SS diquarks from color octet exchange an altogether non-negligible contribution to net baryon density at mid-rapidity.

Account taken that the presently studied sample of "multiple collision" pC reactions corresponds to an average number of ~2.6 carbon nucleons hit by the projectile[2], it is already maximally sensitive to the very interesting contribution from the no diquark (0 type) diagram as evident from Table 1. Thus the very small size of this contribution, suggested by the data and apparent in Fig. 1, does not speak optimistically for a possible experimental identification of baryons from this diagram in the foreseeable future.

The study presented in Fig. 1 shows that already for the collision of the proton with the relatively small C nucleus, an essential role is played by new color configurations involving sea quarks (VS and SS type), not available in pp collisions. The presence of these new configurations explains the earlier failure of valence-diquark preserving calculations to explain the experimental data [1].

---

[2] For more details, see Refs. [1] and [9].

## 8. Baryon stopping as a function of the number of collisions

In this section we address the consequences of the findings made in this letter which we quantify in terms of calculations for net proton spectra as a function of the number of nucleons hit by the projectile proton which corresponds to the number of exchanged soft gluons. For the present calculation we approximate the GEM model by its dominant contribution which is color octet exchange (the $R_8^{N-1}$ representation). The justification can be found in Fig. 3 which shows a consistently good description of the data by the model.

The (absolutely normalized) proton rapidity distributions obtained by GEM for the case of N=2, 3, 4 and 6 collisions are shown in Figs. 4 a, b, c and d respectively. In order to allow for a better inspection of the increase of nuclear stopping power, the proton distribution from NA49 data at ~2.6 collisions is included in each plot. An evident trend emerges from the four Figures: the contribution from the valence diquark (VV type), dominant at N=2, rapidly decreases with N, while at the same time the far more central sea-sea configuration (SS type), non-existent at N=2, emerges at N=3 and rapidly increases with N. With the valence-sea (VS) configuration being reasonably stable as a function of N, this results in a rapid "push" of initial baryon number towards lower rapidities (nuclear stopping power increasing as a function of the number of collisions). Thus the origin of the latter increase of nuclear stopping power lies in the emergence of new (VS, SS) color configurations as a function of the number of exchanged color octets rather than in the energy loss of the original valence (VV) diquark.

## 9. Summary and discussion

In this letter being a direct continuation of our letter [1] we applied the Gluon Exchange Model introduced therein to pA reactions involving N>1 proton-nucleon collisions. On the basis of a simple assumption of a statistical, *i.e.* equally probable formation of diquarks from valence and sea quarks in a color antitriplet state, we developed a full algebra for the emergence of new color configurations as a function of the number N of exchanged gluons. The only "free" parameter of our model, that is the contribution of the color decuplet exchange to the baryon stopping process, appeared very strongly constrained and limited by the experimental data. Our application of color algebra to the multiple collision process brings a satisfactory description of the very precise and complete experimental information on proton and neutron emission at the CERN SPS [2,3].

The "minimalistic" - or simply rigorous - character of our model, involving only a specific color algebra apart from elements developed earlier for pp collisions in Ref. [1], puts it in our mind in strong contrast to the earlier ideas by Capella and Salgado [10,11]. Contrary to our present work, the latter included an additional dynamical element (the string junction based on the concept by Rossi and Veneziano [12]). The quantitative constraints resulting from color algebra were, to the best of our knowledge, not included explicitly. The subdivision of the baryon spectrum into diquark-preserving (DP) and diquark-breaking (DB) components made in the cited works brings some degree of similarity with respect to the presence of VV and VS configurations resulting from our color algebra but the SS component, essential for the description of pA collisions at high N (atomic mass, centrality) is completely absent in Refs. [10,11]. Also the "no diquark" configuration resulting from color decuplet exchange is missing in the cited works. A more detailed comparison of the two approaches remains beyond our present scope because it would necessitate a reanalysis of the new, very restrictive data [2,3] (including not only protons but also neutrons which is not characteristic of the earlier datasets) by the earlier model [10,11] in order to further verify its validity.

The picture of nuclear stopping power resulting from our analysis shows the baryon stopping process as resulting from color (soft gluon) exchange, and governed by the emergence of new color configurations involving valence and sea quarks. These configurations depend on the number of exchanged gluons and therefore are richer in the multiple collision process which is the driving force for the increase of nuclear stopping power as a function of nuclear size, or centrality of the pA reaction. In particular, effective diquarks involving only sea quarks (SS type) appear essential in the process.

The fact that baryon spectra are governed by the emergence of new color configurations, not associated with the original valence diquark and not available in pp collisions, explains the failure of valence diquark-preserving models in explaining the nuclear stopping power in our recent works [1] as well as in the past [13]. Importantly, as a consequence the baryon stopping process appears to be very strongly connected to the number of proton-nucleon collisions but much less, if at all, to the energy loss of the original valence diquark.

**Acknowledgments**

This work was supported by the National Science Centre, Poland (grant no. 2014/14/E/ST2/00018).

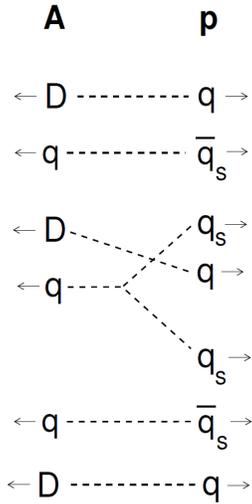

Fig. 1. The basic diagram for color singlet (string) formation for the color configuration of the SS type as described in the text, drawn for the case N=3.

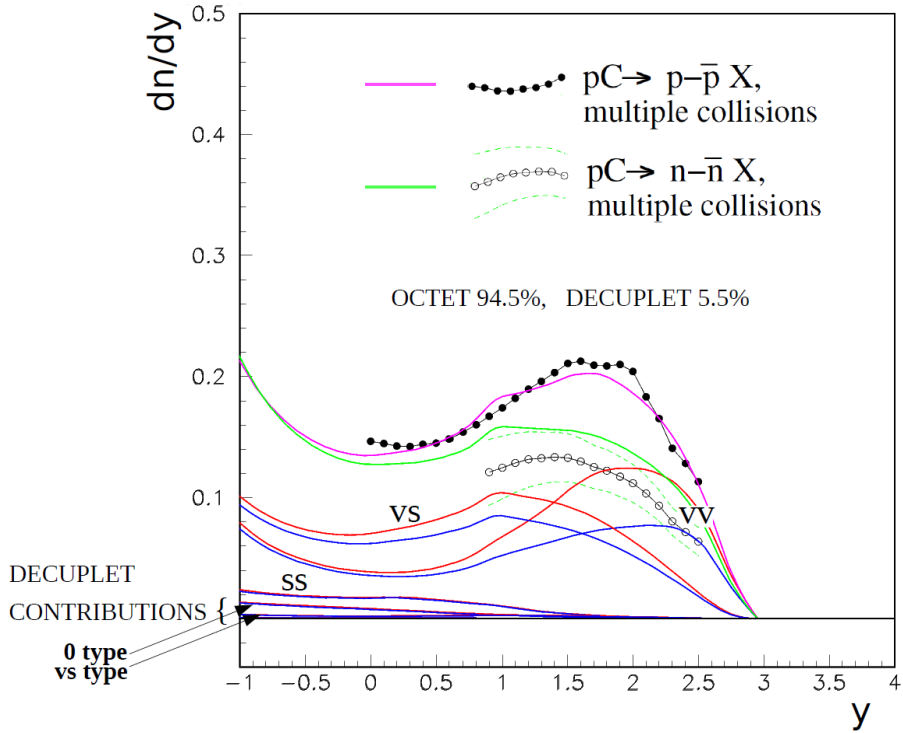

Fig. 2. Rapidity distribution of net protons and net neutrons in pC reactions in which the projectile proton undergoes more than one collision with carbon target nucleons, obtained from the NA49 experiment [2,3], and compared to our GEM calculation described in the text. The collision energy is $\sqrt{s_{NN}}$=17.3 GeV. The calculation for protons (neutrons) is drawn in magenta (green) for the total distribution and in red (blue) for the contributions. The dominant contributions from color octet exchange are indicated as VV, VS and SS type. For the small contributions from color decuplet exchange only the 0 and VS type are indicated by arrows as the SS contribution is too small to be visible in the plot. Note: for the SS configuration from color octet exchange and all the decuplet contributions, the difference between protons (red) and neutrons (blue) remains nearly invisible in the plot.

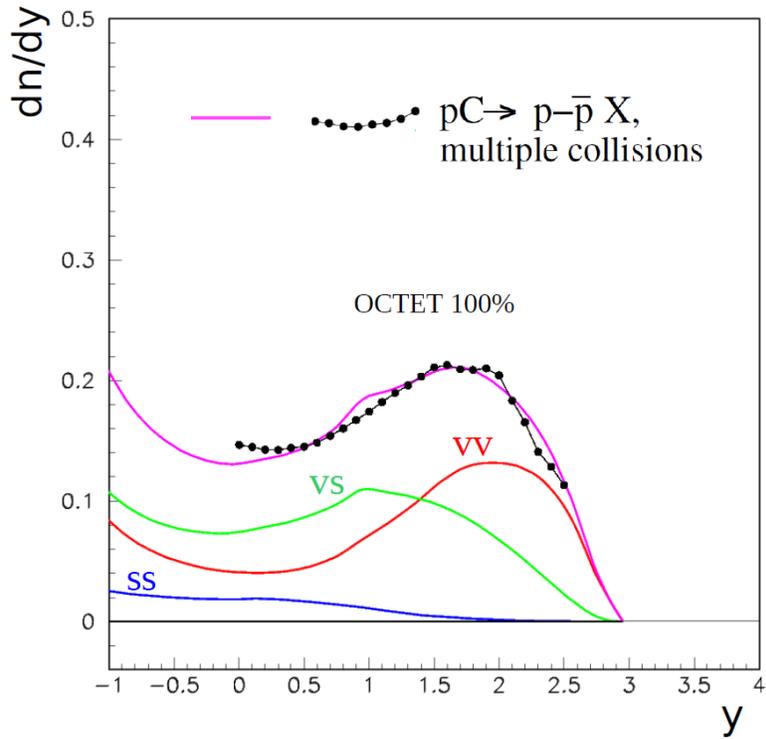

Fig. 3. Rapidity distribution of net protons in pC reactions in which the projectile proton undergoes more than one collision with carbon target nucleons, obtained from the NA49 experiment [2,3], and compared to our GEM calculation assuming 100% color octet exchange. The contributions from the VV, VS and SS configurations are indicated in the plot.

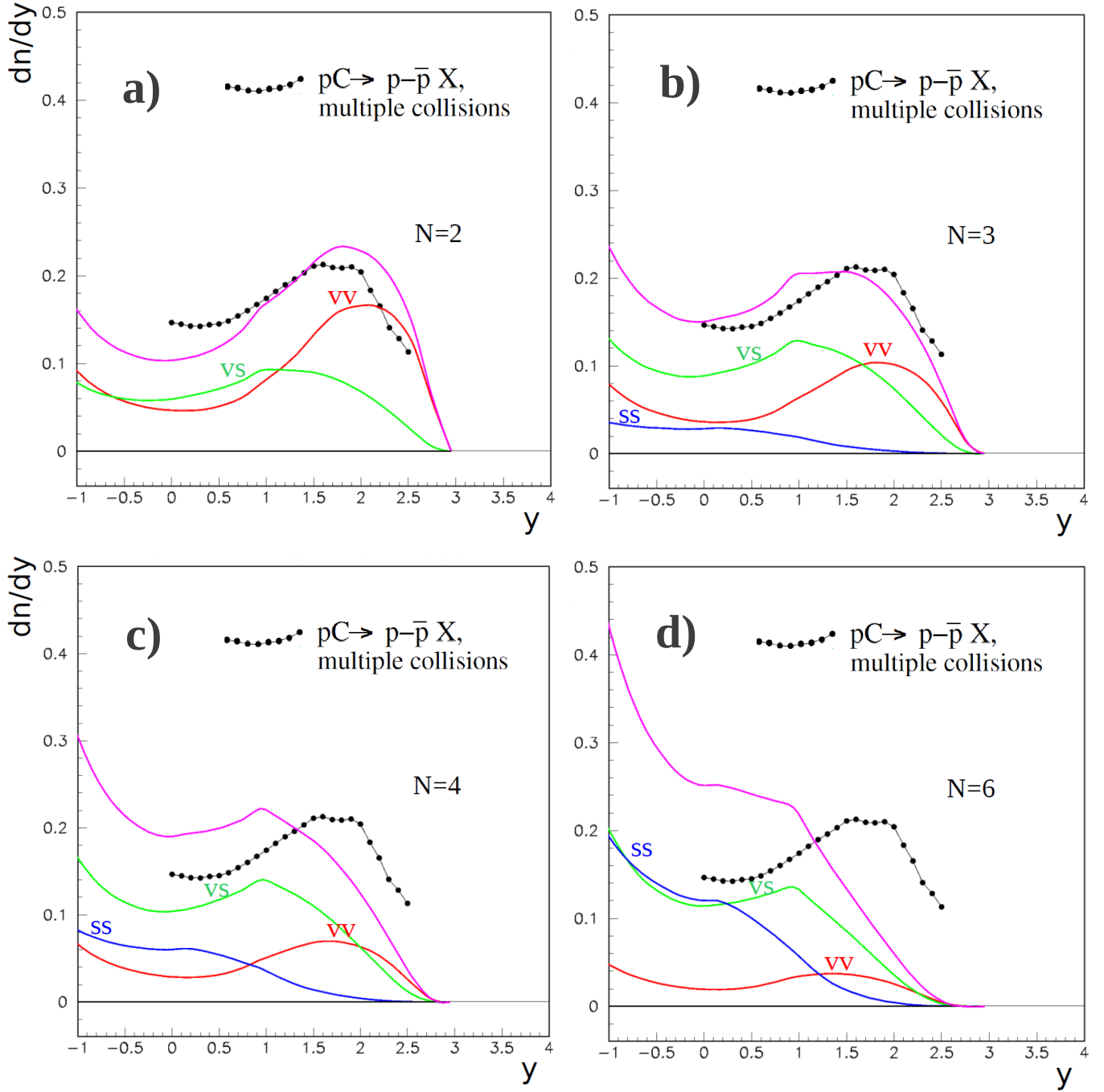

Fig. 4. Rapidity distributions of net protons obtained by the GEM model assuming 100% color octet exchange, for pA reactions with N=2, 3, 4 and 6 nucleons hit by the projectile proton. The contributions from the VV, VS and SS color configurations are indicated. In order to allow for a better inspection of the increase of nuclear stopping power, the proton distribution obtained from NA49 data [2,3] is included in each plot.